\documentclass[superscriptaddress,reprint,nofootinbib,preprintnumbers,amsmath,amssymb,aps]{revtex4-2}
\usepackage{amsmath, mathrsfs, latexsym, amssymb, color, graphicx, comment, hyperref, scrhack}
\usepackage{bbm}
\usepackage{lmodern}
\usepackage{physics}

\usepackage{graphicx}
\usepackage{color}
\usepackage[usenames,dvipsnames]{xcolor}

\begin{document}

\title{
$\mathbbm{Z}_2$ gauge and exact solution of the $\pm J$ Edwards–Anderson model in a transverse field
}
\author{M.\,G.\,Vasin}

\begin{abstract}
Using the Majorana representation of spin-1/2 operators, we study the $\pm J$ Edwards--Anderson model in a transverse field at $T=0$. We show that, in this representation, the quantity describing the coupling between neighboring spins commutes with the Hamiltonian and, unlike in the Kitaev model, is not subject to any Gauss law. The resulting $\mathbbm{Z}_2$ gauge freedom allows the disorder in the signs of the exchange integrals to be eliminated by an exact gauge transformation, so that the spectrum of the system coincides with that of the model with uniform couplings. The resulting theory is quadratic in the Majorana fields. Computing the local Green's functions, we find that the Edwards--Anderson order parameter vanishes for $\Delta>J\sqrt{Z}$ and equals $1/4$ for $\Delta<J\sqrt{Z}$, which signals a second-order transition from a paramagnetic phase to a spin-glass ground state that is degenerate over spin configurations and replica symmetric at $T=0$.
\end{abstract}

\maketitle


The Edwards--Anderson (EA) model \cite{EA} is one of the fundamental models of condensed matter physics, describing spin glasses with short-range interactions. Its Hamiltonian has the form
\[
H = -\sum_{\langle ij\rangle} J_{ij} \sigma^x_i \sigma^x_j - \Delta \sum_i \sigma^z_i,
\]
where $J_{ij} = \pm J$ are random exchange integrals (binary distribution), $\Delta$ is the transverse magnetic field, and $\sigma^l_i$ are Pauli matrices. In the classical limit ($\Delta = 0$), this model exhibits a rich ground-state structure with many local minima separated by high energy barriers. The random sign of $J_{ij}$ leads to frustration --- a situation in which it is impossible to simultaneously satisfy all interactions, which gives rise to degeneracy and complex landscape dynamics \cite{Mezard}.
Adding a transverse magnetic field to the Edwards--Anderson model at absolute zero temperature converts the problem from classical to quantum.
The transverse field $\Delta$ plays the role of a tunneling term that allows the system to overcome barriers between local minima.
Now the spins cannot only ``freeze'' in one of the many classical configurations but also quantum-tunnel between them. This fundamentally changes the physics of the system.
For this reason, the quantum version of the EA model attracts particular attention in the context of quantum annealing --- a method used in quantum computing to solve complex optimization problems \cite{Kadowaki,Nishimori}.
However, the theoretical description of the critical behavior of such a system remains a difficult problem due to the simultaneous presence of disorder and quantum fluctuations.

In recent years, significant progress in understanding quantum spin systems has been achieved through the use of fermionic representations. A special place is occupied by the approach proposed by Kitaev \cite{Kitaev2006}, in which spin-1/2 degrees of freedom are mapped onto Majorana fermions. This representation made it possible to exactly solve a number of models, including the famous Kitaev model on the honeycomb lattice, which exhibits a topological phase with non-Abelian statistics.
In this work we apply the method proposed by Kitaev: we apply the Majorana representation of spins to the binary EA model and discover a remarkable property: the operators $u_{ij} = \mathrm{i}\psi^z_i \psi^z_j$, defined on each bond, commute with the Hamiltonian. This allows them to be used as gauge degrees of freedom.

Unlike the Kitaev model, where analogous operators are subject to topological constraints (the Gauss law for a $\mathbbm{Z}_2$ gauge field), in our case the operators $u_{ij}$ are formed only by the $z$-components of a pair of spinors. These $z$-type bonds form a set of independent dimers and cannot form closed loops. Consequently, the operators $u_{ij}$ can be chosen independently for each bond, which makes it possible to flip the sign of the random $J_{ij}$ and eliminate frustration. The original problem of the model with random signs of couplings reduces to the problem with positive coupling constants $J=|J_{ij}|$. This cardinal simplification allows the quantum paramagnet--spin-glass transition to be described as a second-order phase transition. We compute the correlation functions of the Majorana fields, and analyze the behavior of the system near the phase transition.

In our model we use the Tsvelik SO(3) representation of spins by Majorana spinors \cite{Tsvelik1993,PhysRevB.97.115142}. To this end, at each lattice site $i$ we introduce three Majorana operators $\psi^x_i, \psi^y_i, \psi^z_i$ \cite{Tsvelik1993} satisfying the Clifford algebra:
\[
\{\psi^l_i, \psi^m_j\} = 2\delta_{ij}\delta^{lm}, \qquad (\psi^l_i)^2 = 1, \qquad \psi^{l\dagger}_i = \psi^l_i.
\]
The spin-1/2 operators are mapped onto these fermions as follows:
\begin{equation}
\sigma^{l}_i = -\frac{\mathrm{i}}{2}\varepsilon^{lmn}\psi^{m}_i\psi^{n}_i, \qquad l,m,n \in \{x,y,z\}.
\label{spin_majorana}
\end{equation}
A direct check shows that this mapping preserves the Pauli algebra: $[\sigma^l_i, \sigma^m_j] = 2\mathrm{i}\varepsilon^{lmn}\delta_{ij}\sigma^n_i$.
In addition, it is important to note that the three Majorana spinors are constrained: $\psi_i^x\psi_i^y\psi_i^z=\pm1$.
Substituting (\ref{spin_majorana}) into the EA Hamiltonian, we obtain:
\begin{equation}
H = -\sum_{\langle ij\rangle} J_{ij} \psi^y_i \psi^y_j \psi^z_i \psi^z_j + \mathrm{i}\Delta \sum_i \psi^x_i \psi^y_i.
\label{H_majorana_full}
\end{equation}
Here, the first term comes from $\sigma^x_i\sigma^x_j$, the second from $\sigma^z_i$. It is important to note that the $z$-components of the Majorana fields enter only through the product $\psi^z_i\psi^z_j$ on each bond.

As in the Kitaev model \cite{Kitaev2006}, we define on each bond $\langle ij\rangle$ an antisymmetric operator
\begin{equation}
u_{ij} \equiv \mathrm{i}\psi^z_i \psi^z_j.
\label{u_operator}
\end{equation}
From the Clifford algebra it follows:
\(
u_{ij}^2 = -\psi^z_i \psi^z_j \psi^z_i \psi^z_j = \psi^z_i \psi^z_i \psi^z_j \psi^z_j = \mathbbm{1} ,
\)
so the eigenvalues of this operator are known up to a sign in any eigenstate: $u_{ij} = -u_{ji} =\pm 1$.
The key property of these operators is:
\begin{equation}
[u_{ij}, H] = 0,
\label{commutation}
\end{equation}
which allows $H$ and all $u_{ij}$ to be diagonalized simultaneously.
In our model, the $z$-component of the Majorana operators enters only through the product $\psi^z_i\psi^z_j$ on a given bond, since $u_{ij}$ contains only $\psi^z_i$ and $\psi^z_j$, while all other terms of the Hamiltonian involve $x$- and $y$-components that anticommute with $\psi^z$; moreover, each site belongs to only one bond with a $z$-coupling, so there are no cross commutators. After replacing $u_{ij} = \pm 1$, the quantity $\psi^z_i\psi^z_j$ disappears completely from the Hamiltonian, and two Majorana components $\psi^x_i, \psi^y_i$ remain at each site, which is equivalent to one complex fermion and coincides with the physical spin space. This property fundamentally distinguishes our model from the Kitaev model \cite{Kitaev2006}: there, the $z$-bonds close into plaquettes and yield the Gauss law $\prod u_{ij} = \pm 1$, whereas in our case, the $z$-bonds form disconnected dimers (each bond is a separate dimer), and there is no global constraint on the set of $u_{ij}$. Therefore, the choice of $u_{ij}$ is independent for each bond, which leads to the irrelevance of disorder in the signs of the couplings for the model under consideration.
Since the field \(u_{ij}\) is free, we can choose the gauge
\begin{equation}
u_{ij} =\operatorname{sgn}(J_{ij}),
\label{gauge_choice}
\end{equation}
which minimizes the Hamiltonian (\ref{H_majorana_full}), and the latter takes the form
\begin{equation}
H = -\mathrm{i} \sum_{\langle ij\rangle} J\varepsilon_{ij}\psi^y_i \psi^y_j + \mathrm{i}\Delta \sum_i \psi^x_i \psi^y_i\quad (\varepsilon_{ij}=-\varepsilon_{ji}=1).
\label{H_uniform}
\end{equation}
Since the operators $u_{ij}$ on different bonds also commute and are independent, the spectrum of $H$ depends on $J_{ij}$ only through the product $J_{ij}u_{ij}$. Therefore, for any configuration of signs $\{J_{ij}\}$, the choice $u_{ij} = \operatorname{sgn}(J_{ij})$ on each bond gives $J_{ij}u_{ij} = J\varepsilon_{ij}$, so that the spectrum coincides with that of the model with uniform couplings $|J_{ij}|$. This is the reason for the irrelevance of sign disorder---it is eliminated by choosing the conserved $\mathbbm{Z}_2$ gauge field $u_{ij}$.

It is important to emphasize that this is not an approximation but an exact transformation. The degeneracy associated with different configurations of the signs $J_{ij}$ is eliminated by the $\mathbbm{Z}_2$ gauge. This makes sign disorder irrelevant for thermodynamics and ensures ergodicity in the quantum sense at the level of gauge sectors. At the same time, within each sector of $u_{ij}$ the degeneracy of spin configurations is preserved: different spin configurations have the same energy. If replicas are associated with these configurations, such a picture indicates the existence at $T=0$ of a replica-symmetric ground state in quantum spin-glass models \cite{PhysRevB.39.11828,Thirumalai_1989,PhysRevB.52.384}.

Remarkably, in this case, it is completely irrelevant on which lattice the model is built, i.e., with which sites and how any chosen site is connected.
It turns out that the spin system in the quantum representation, owing to the larger number of degrees of freedom, ``bypasses'' the energy barriers that are insurmountable for systems of classical spins, which makes the theory exactly solvable (the correlation functions can be computed).

To analyze the critical behavior, we use the formalism of time-dependent Green's functions. We write the Lagrangian corresponding to the Hamiltonian (\ref{H_uniform}) of the binary model ($J=|J_{ij}|$):
\begin{equation}
\mathcal{L} = -\mathrm{i} \sum_{\langle ij\rangle}J\varepsilon_{ij}\psi^y_i\psi^y_j - \sum_i \left(\sum_{a=x,y,z} \psi^a_i \partial_t \psi^a_i + \mathrm{i}\Delta \psi^x_i\psi^y_i\right).
\label{Lagrangian}
\end{equation}
From this expression, it is clear that the action is quadratic in the spinor fields.
It is not difficult to obtain the bare local correlators:
\(
\langle \psi^a_i\psi^b_i\rangle_0 = \mathrm{i}\delta^{ab}({\omega + \mathrm{i}\epsilon})^{-1},
\)
where $\epsilon \to 0^+$ is an infinitesimal imaginary part ensuring the correct pole prescription.
\begin{figure}[h!]
    \centering 
    \includegraphics[scale=0.5]{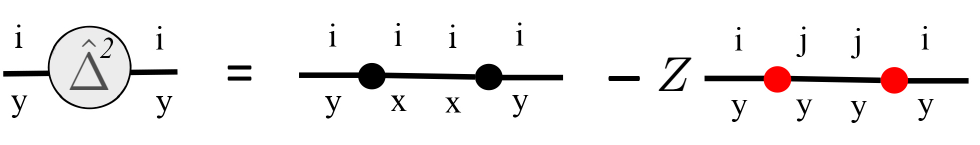}
    \caption{Graphical representation of the self-energy entering the correlation function \(\langle \psi^y_i\psi^y_i\rangle\). The line corresponds to the bare correlator \(\langle \psi^{\alpha}_i\psi^{\alpha}_i\rangle_0\). The black dot corresponds to the $\Delta$ vertex, the red (grey) dot to $J$, corresponding to a hop between lattice sites. The factor \(Z\) in front of the second term means that the trajectory can leave site \(i\) for any of the \(Z\) sites and return immediately. This term represents symmetric hopping (two-site hopping). The minus sign in front of it arises because $u_{ij}=-u_{ji}$ and the commutation relation $[\sigma^z_i\sigma^z_j]=0$ holds for $i\neq j$.}
    \label{fig1}
\end{figure}
The interaction between the $x$- and $y$-components arises due to the term $\mathrm{i}\Delta \psi^x_i\psi^y_i$. In addition, for the $y$-components, one can write an additional interaction arising from symmetric hopping (two-site hopping) with neighboring sites (Fig.\,\ref{fig1}). Thus, the local correlators have the form:
\begin{gather}
\langle \psi^y_i\psi^y_i\rangle_{\omega} = \frac{\mathrm{i}\omega}{\omega^2 + \Delta^2 - ZJ^2},
\
\langle \psi^x_i\psi^x_i\rangle_{\omega} = \frac{\mathrm{i}\omega}{\omega^2 + \Delta^2},
\\
\langle \psi^z_i\psi^z_i\rangle_{\omega} = \frac{\mathrm{i}}{\omega + \mathrm{i}\epsilon},
\
\langle \psi^x_i\psi^y_i\rangle_{\omega} = \frac{\mathrm{i}\Delta}{\omega^2 + \Delta^2}.
\label{corr_xy}
\end{gather}
Here $Z$ is the coordination number of the lattice (the number of bonds connecting a given spin to other spins).
From (\ref{Lagrangian}) it is clear that collective effects are determined by the correlator $\langle\psi^y_i\psi^y_i\rangle$, whose pole corresponds to the effective mode
\begin{gather}
\hat\Delta \equiv \sqrt{\Delta^2 - J^2Z},
\label{eff_gap}
\end{gather}
of the collective excitation of the entire spin ensemble.
Thus, at $\Delta=\Delta_c\equiv J\sqrt{Z}$ a second-order phase transition occurs in the system.
To describe it, it is convenient to use the Edwards--Anderson parameter as a universal order parameter:
\begin{multline}
q_{EA}=\frac1N\sum^N_{i=1}\langle\sigma^x_i\sigma^x_i\rangle_{t\to \infty}=-\frac1{4N}\sum^N_{i=1}\langle \psi^y_i \psi^z_i,\psi^y_i \psi^z_i\rangle_{t\to \infty}\\=
\begin{cases}
0, & \Delta>J\sqrt{Z},\\[6pt]
\frac14, & \Delta<J\sqrt{Z}.
\end{cases}
\label{sigma}
\end{multline}
Such a change of the order parameter indicates that a phase transition occurs from a paramagnetic phase to a spin-glass phase characterized by the existence of two peaks $q_{EA}=\pm 1/4$ in the distribution function of the order parameter $P(q_{EA})$. This is consistent with the results of calculations of the quantum model, in which (unlike the classical case) these two peaks dominate as the size of the simulated system increases \cite{PhysRevE.110.065305}, and indicates the existence in the spin-glass phase of a ground state degenerate over spin configurations.

Usually, two-dimensional models are considered on a square lattice, for which $Z=4$. Substituting this value, we find $\Delta_c\approx 2J$, which also agrees well with the results of numerical calculations \cite{King2023,Bernaschi2024,PhysRevE.110.065305}. At the same time, it is important to recall that we took into account symmetric hopping only between nearest neighbors, and to refine the estimate one should include higher-order terms of the expansion.

As another example, let us consider a one-dimensional spin chain whose Hamiltonian is written as
\[
H = -\sum_{i} J_{ij} \sigma^x_i \sigma^x_{i+1} - \Delta \sum_i \sigma^z_i.
\]
According to the above, the Lagrangian of the binary model can be represented in the following form:
\begin{gather*}
\mathcal{L} = -\mathrm{i}J \sum_{i} \psi^y_i\psi^y_{i+1} - \sum_i \left(\sum_{a=x,y,z} \psi^a_i \partial_t \psi^a_i + \mathrm{i}\Delta \psi^x_i\psi^y_i\right).
\label{Lagrangian_1}
\end{gather*}
It is important to note that in the one-dimensional case $Z=2$, but unlike multidimensional lattices, in the case of a one-dimensional chain one must take into account reverse hopping in only one direction, since backward hops are already included in the renormalization of the previous site. Therefore the effective gap (\ref{eff_gap}) has the form $\hat\Delta \equiv \Delta - J$. The phase transition in the system occurs when this gap closes, i.e., at $\Delta=\Delta_c\equiv J$.

We obtain the known result \cite{Nishimori,10.1093/acprof:oso/9780198509233.001.0001, PhysRevB.51.6411} --- the critical behavior of Ising spin chains with constant and random couplings is the same. Indeed, even in the classical pure 1D Edwards--Anderson model, disorder does not change the critical point, since the Hamiltonian with an arbitrary distribution of couplings $J_{ij}$ (including random $\pm J$) is gauge-equivalent to the Hamiltonian with uniform couplings. A sign flip of a coupling at one site can be ``compensated'' by flipping the corresponding spin (a gauge transformation), so disorder does not create new frustrations that would destroy long-range order \cite{suzuki2013quantum}.

We note that the spin-1/2 chain in the TFIM, after the Jordan--Wigner transformation, maps onto the Kitaev model, which is the canonical model for realizing topological protection of Majorana edge modes,
and the quantum phase transition described above is a transition between the ``trivial'' ($\Delta>J$) and ``topological'' ($\Delta<J$) phases. Indeed, as was shown above, introducing any local perturbations that arbitrarily change the sign of couplings does not affect the ground state of the topological phase.
However, introducing a perturbation that breaks the $\mathbbm{Z}_2$ symmetry into the system leads to the disappearance of topological protection.

The results presented above allow us to
answer the still-open question of whether the spin-glass state in the quantum model is replica-symmetric or non-ergodic, as in the classical case \cite{suzuki2013quantum}. In addition, these results are of fundamental importance for quantum computing via quantum annealing, for which the Edwards--Anderson model in a transverse field is a paradigmatic hard optimization problem and provides a solid theoretical basis for understanding and demonstrating quantum critical dynamics. The results are directly relevant to the problem of the efficiency of quantum annealing: the relaxation time of the system near the transition determines the rate of adiabatic evolution and, consequently, limits the performance of quantum computing devices. The analytical solution presented in this work allows us to assert that in the considered Edwards--Anderson model in a transverse field at $T=0$, quantum annealing brings the system directly to the ground state. For this reason, spin-glass effects associated with the frustration of the system and leading to a significant slowdown of this process are absent.


\bibliography{PRL.bib}

\end{document}